\documentclass[usenatbib]{mn2e}
\usepackage{times}
\usepackage{epsfig}
\newcommand{\be}{\begin{equation}}
\newcommand{\ee}{\end{equation}}
\newcommand{\etal}{et al.}
\newcommand{\dev}{de Vaucouleurs}

\newcommand{\labsecn}[1]{\label{sec:#1}}

\newcommand{\kms}{~km~s$^{-1}$}

\begin{document}
\title[Scaling relations in early type galaxies]{Scaling relations in
     early-type galaxies belonging to groups}
\author[Habib Khosroshahi et al.]{ %
       \parbox[t]{\textwidth}{Habib G. Khosroshahi$^{1}$\thanks{E-mail:
habib@star.sr.bham.ac.uk}, Somak Raychaudhury$^{1}$,
Trevor~J.~Ponman$^{1}$,
Trevor~A.~Miles$^{1}$ \&
Duncan~A.~Forbes$^{2}$}
\vspace*{6pt} \\
$^{1}$School of Physics and Astronomy, The University of Birmingham,
Birmingham B15 2TT, UK\\
$^{2}$Centre for Astrophysics and Supercomputing, Swinburne University,
Hawthorn, VIC 3122, Australia}
\date{
MNRAS submitted - September 2003
}
\pagerange{\pageref{firstpage}--\pageref{lastpage}}
\pubyear{2003}
\label{firstpage}
\maketitle
%
\begin{abstract} 
We present a photometric analysis of a large sample of early-type
galaxies in 16 nearby groups, imaged with the Wide-Field Camera on the
Isaac Newton Telescope.  Using a two-dimensional surface brightness
decomposition routine, we fit Sersic ($r^{1/n}$) and exponential
models to their bulge and disk components respectively. Dividing the
galaxies into three subsamples according to the X-ray luminosities of
their parent groups, we compare their photometric properties.
Galaxies in X-ray luminous groups tend to be larger
and more luminous than
those in groups with undetected or low X-ray luminosities, but no
significant differences in $n$ are seen.  Both normal and dwarf
elliptical galaxies in the central regions of groups are found to have
cuspier profiles than their counterparts in group outskirts.

Structural differences between dwarf and normal elliptical galaxies
are apparent in terms of an offset between their ``Photometric
Planes'' in the space of $n$, $r_e$ and $\mu_0$. Dwarf ellipticals are
found to populate a surface, with remarkably low scatter, in this
space with significant curvature, somewhat similar to the surfaces of
constant entropy proposed by
\citet{mar01}.  Normal ellipticals are offset from this distribution
in a direction of higher specific entropy. This may indicate that the
two populations are distinguished by the action of galaxy merging on
larger galaxies.

\end{abstract}


\begin{keywords}
galaxies: fundamental parameters --- galaxies: evolution ---
galaxies: dwarf ---
galaxies: elliptical and lenticular, CD ---
galaxies: structure --- galaxies: groups --- X-ray: galaxies: clusters
\end{keywords}

\section{Introduction}

Galaxy scaling relations, such as the Fundamental Plane, and its
various projections, and the colour-magnitude relation have been used
to enhance our understanding of galaxy structure and
evolution. Early-type galaxies, in particular, form a relatively
homogeneous population and hence rather well-defined scaling
relations. Most large samples of early-type galaxies come from the
rich environments of galaxies clusters. Studies of these high density
regions benefit from the large numbers of galaxies in a small angular
region of the sky (hence making observations more efficient) and from
the morphology-density relation, which tells us that such environments
are dominated by early-type galaxies. Thus our current knowledge
gained from galaxy scaling relations applies mostly to clusters.  The
field and group environments are less well studied in this respect,
and provide a means to study environmentally-dependent processes. For
example, galaxy mergers (which can be about hundred times more
efficient in today's groups than clusters, cf.
\citealt{mamon86}) may result in merger-remnant Ellipticals that
deviate strongly from the scaling relations \citep[e.g.][]{fpb98}.
On the other hand many of the well studied Elliptical galaxies are
brightest group galaxies.

Recent X-ray observations of groups and their constituent galaxies
have provided a new insight into the study of the evolution of
galaxies in small ensembles and in intermediate-density
environments \citep{hp00,mul00}. \citet{miles03} find that the
shape of the luminosity function of galaxies in groups with low
X-ray luminosity is significantly bimodal, in comparison with that
of X-ray bright groups or rich clusters, showing a strong
environmental effect on the morphological mix
in a system, and on the evolution of galaxies in general. It is likely
that the dip in the luminosity function at intermediate magnitudes
arises from increased merging in low velocity dispersion systems.

Of particular interest in this context are early-type galaxies, which
are more abundant in clusters and groups than in the field.  Several
well-known scaling relations (such as the Faber-Jackson
relation or the Fundamental Plane of Elliptical galaxies) utilise
galaxy velocity dispersions and hence require spectroscopic data.
However, scaling relations like the Kormendy relation \citep{korm77}
and the Photometric Plane of galaxies
\citep{habib00a} can be constructed from photometric observations
alone.

There have been many studies
aimed at understanding the differences in the structure of
Ellipticals (E) and dwarf Ellipticals (dE), with mixed results.
While some \citep[e.g.][]{ter01} argue in favour of the
similarities between the two populations, others 
\citep[e.g.][]{korm85} find evidence otherwise (see \citealt{graham03} 
for more details). 

Traditionally, the surface brightness distribution of Ellipticals and
bulges of disk galaxies have been modelled by the \dev\
profile. However, during the last decade, better photometry and
advanced analysis have shown that their surface brightness
distribution is, in general, better represented by the Sersic profile
($r^{1/n}$,
\citealt{sersic68}), of which the \dev\ profile ($r^{1/4}$) is a
special case \citep{caon93,habib00a,graham01}. The Sersic model is
also widely used to describe the brightness distribution of bulges of
disk galaxies
\citep{dejong96,and95,habib00b}. The disk component of galaxies is
usually characterised by an exponential distribution.

In this paper we examine the surface brightness distribution of
Ellipticals belonging to galaxy groups, where most galaxies in the
Universe reside.  The relatively small number of galaxies in
individual groups, compared to clusters, requires any analysis to be
carried out on a large sample. This can be done in several ways.
Here, the galaxies are classified based on the X-ray luminosity ($L_X$) of
their parent groups, which, being correlated with the velocity
dispersion of the group \citep[e.g.][]{hp00}, can be an index of
the dynamical state and the mass of the group
\citep{miles03,osmond03}. The principal properties of
the groups, and a detailed description of the sample can be found in
section 2. The analysis techniques and morphology of galaxies is
discussed in section 3. Correlations among various parameters and the
scaling relations is studied in section 4. Section 5 contains a
discussion and a summary of results.

\section{Observations and Analysis}

\subsection{Observations}

Our sample is drawn from the Group Evolution Multi-wavelength
Study \citep[GEMS,][]{osmond03} of sixty groups, compiled by
cross-correlating a list of over 4000 known groups from various
studies with the list of archival ROSAT PSPC X-ray observations
with integration $>10$~ks. This includes several Hickson groups,
as well as loose groups from the CfA survey. A large fraction of
these were detected in the X-ray, and for the others we have upper
limits for their X-ray luminosity.

Of the GEMS sample, 16 groups were observed at the 2.5m Isaac
Newton telescope at the Roque de Los Muchachos Observatory, La
Palma, between 2000 February 4--10.  This is a random and 
representative subsample of the original sample of 60 groups, 
where all groups accessible for the allocated observing run 
were observed. The detector used was the
Wide Field Camera (WFC), which is an array of four thinned EEV
CCDs situated at the f/3.29 prime focus of the INT, each with an
area of $2048\times 4096$ pixels, each pixel being 0.33~arcsec
across. Each CCD thus can image an area of $22.5\times 11.3$
arcmin of sky, together covering 1017 sq.arcmin.

Photometry was performed with broadband BRI filters, of which
we analyse only the $R$-band images here. Our analysis is limited 
to galaxies brighter than $M_R=-13$. The median seeing achieved was about 
1.1 arcsec in $R$. Further details can be found in \citep{miles03}.

\subsection{Group Membership}
\label{colourselection}

The identification of the group galaxies is based on a colour
selection. Source extraction was performed using the Sextractor
software \citet{sext}, which  uses a routine based on neural
network generated weights for star-galaxy separation.  Detections
were checked visually and objects with the {\em stellaricity}
parameter greater then 0.9 were deemed to be definitely stellar
and therefore discarded. To obtain colours, a fixed aperture, set
to be slightly greater than the seeing, was used to evaluate
fluxes with different filters.  All objects with FWHM less than
the mean seeing were not processed further.

Galaxies were selected as being likely group members on the basis
of their $B\!-\!R$ colour.  A conservative cut of $B\!-\!R\! =\!
1.7$ was found to remove the majority of background galaxies for
all groups in our sample (See \citealt{miles03} for  further
details). Those galaxies found in the NASA/IPAC Extragalactic
Database (NED) within the virial radius $R_{\rm vir}$ 
\citep[see][]{osmond03}.

\begin{table}
\begin{center}
\caption{The sample of 16 groups used in this study
\label{table1}
}
\begin{tabular}{lrrrcr}
\hline
Group & R.A.  &  Dec. & N$^a$
&   Distance$^b$  &$\log~L_X\>^c$   \\
& (J2000)  & (J2000) &  & Mpc  & ergs/s    \\
\hline
HCG   10 & 01:25:40.4 & +34:42:48 & 45 & 68.2   & 41.75    \\
HCG   68 & 13:53:26.7 & +40:16:59 & 177& 41.1   & 41.67    \\
NGC  524 & 01:24:47.8 & +09:32:19 & 27 & 35.4   & 41.09    \\
NGC 1052 & 02:41:04.8 &--08:15:21 & 10 & 20.3   & 41.03    \\
NGC 1587 & 04:30:39.9 & +00:39:43 & 54 & 55.2   & 41.22    \\
NGC 2563 & 08:20:35.7 & +21:04:04 & 149& 73.5   & 42.54    \\
NGC 3227 & 10:23:30.6 & +19:51:54 & 18 & 26.5   & 40.88    \\
NGC 3607 & 11:16:54.7 & +18:03:06 & 26 & 23.5   & 41.09    \\
NGC 3640 & 11:21:06.9 & +03:14:06 & 56 & 28.5   & $<$40.34 \\
NGC 3665 & 11:24:43.4 & +38:45:44 & 41 & 37.2   & 41.15    \\
NGC 4151 & 12:10:32.6 & +39:24:21 & 14 & 23.0   & $<$40.15 \\
NGC 4261 & 12:19:23.2 & +05:49:31 & 43 & 34.8   & 41.84    \\
NGC 4636 & 12:42:50.4 & +02:41:24 & 24 & 13.1   & 41.73    \\
NGC 4725 & 12:50:26.6 & +25:30:06 & 7  & 25.1   & 41.00    \\
NGC 5044 & 13:15:24.0 &--16:23:06 & 55 & 34.4   & 43.08    \\
NGC 5322 & 13:49:15.5 & +60:11:28 & 18 & 32.4   & 41.51    \\
\hline
\end{tabular}
\end{center}
$^a$Number of member galaxies within our imaging field identified
by the colour selection outlined in \S~\ref{colourselection}. \\
$^b$Distance measured from the mean redshift
of group (assuming $H_0=70$ \kms Mpc$^{-1}$; $q_0=0.5$),
corrected for local bulk flows.\\
$^c$Bolometric X-ray luminosity from \citet{osmond03}
\end{table}

\begin{figure}
\epsfig{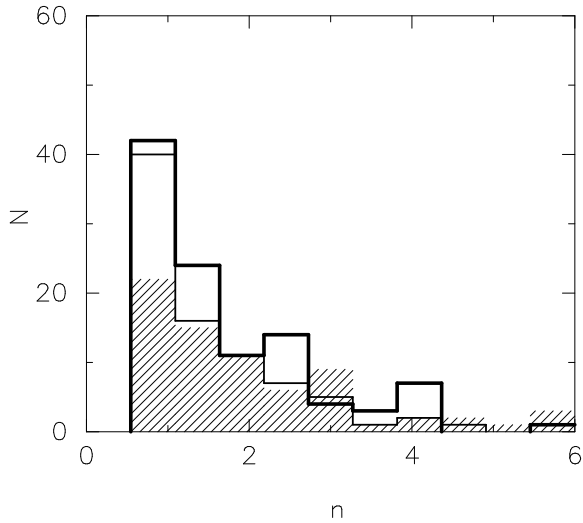} 
\caption{ Distribution of the
Sersic parameter $n$ for early-type galaxies ($B/T>0.5$) 
in our sample,
in X-ray bright (bold line), X-ray dim (thin line) and
X-ray undetected groups (those with no discernible group emission,
shaded histogram). There is
no significant difference in the $n$ distribution between the types of groups.
\label{fig:nhist}}
\end{figure}

\subsection{Analysis}

The core of our data analysis is the two-dimensional decomposition
of galaxies, by fitting a Sersic profile to the bulge and an
exponential profile to the disk component,
for obtaining the structural photometric parameters,
and morphological classification. This analysis was carried
out on all galaxies identified as group members using the selection
method described in the previous section.

The bulge-disk decomposition procedure used here involves fitting
two-dimensional image models to the observed galaxy images, assumed to
consist of a bulge and a disk component.  The model image is convolved
with a Gaussian PSF, with the FWHM measured from the galaxy frame.
Details of the accuracy and reliability of the decomposition
procedure, and the associated galaxy simulation code, can be found in
\citet{wadad99}.

The Sersic profile for the bulge is given by
\begin{eqnarray}
I_{\rm bulge}\,(x,y) &=& I_b(0) \exp\left[
            { -2.303\, b_n (r_{\rm bulge}/r_e)^{1/n}}\right]; \\
r_{\rm bulge} &=& \sqrt{x^2 + y^2/(1 - \epsilon_b)^2}. \nonumber
\end{eqnarray}
Here $x$ and $y$ are the distances from the centre of the galaxy
along the major and minor axis respectively. The quantity $b_n$ is
a function of $n$, and is evaluated as a root of an equation
involving the incomplete gamma function \citep{wadad99}. It is
well approximated by
\begin{equation}
b_n  = 0.87\, n - 0.14. \nonumber
\end{equation}
For $n=4$, which corresponds to the de Vaucouleurs law, $b_4 =
3.33$. We model the disk intensity by the \citet{freem70}
exponential distribution,
\begin{eqnarray}
I_{\rm disk}(x,y) &=& I_d(0) \exp\left({- r_{\rm disk}/r_d}\right);\\
r_{\rm disk} &=& \sqrt{x^2 + y^2/(1 - \epsilon_d)^2}. \nonumber
\end{eqnarray}
The disk is assumed to be intrinsically circular, with the
observed ellipticity $\epsilon_d$ of the disk in the observed
image being due to projection effects alone.

The free parameters of the fit are:
\begin{enumerate}
\item The central
bulge intensity $I_b(0)$,
\item The half-light radius of the bulge $r_e$,
\item The ellipticity of the bulge $\epsilon_b$,
\item The bulge shape parameter $n$ (Sersic profile),
\item The central intensity of the disk $I_d(0)$,
\item The scale length of the disk $r_d$, and
\item The ellipticity of the disk $\epsilon_d$.
\end{enumerate}
Additional parameters such as sky background and position angle
are estimated in a preliminary fit and are then held fixed during
the final fitting process.

\section{Modelling the two-dimensional surface brightness of galaxies}
\labsecn{results}

We were able to obtain satisfactory fits with a reduced
$\chi_{\nu}^{\,2} \le 2$ for 471 galaxies from our initial sample of
764 galaxies.  Any fit with a reduced
$\chi_{\nu}^{\,2} > 2$ is considered
to be unacceptable and the galaxy excluded from the sample. Such
unacceptable fits result mainly from irregularities in the
brightness distribution, or from the presence of multiple nuclei
or close companions. Such features cannot be represented by the simple
and smooth models used in this study.

\begin{figure}
\epsfig{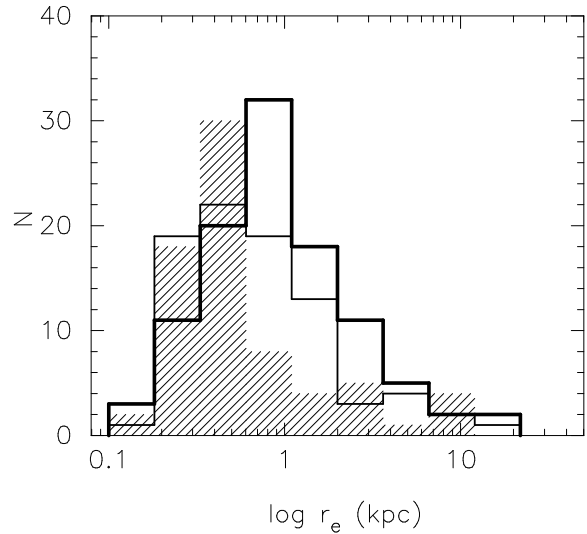}
\caption{Distribution of the half-light radius
$r_e$ for early-type galaxies, 
in X-ray bright (bold line), X-ray dim (thin line) and
X-ray undetected groups (shaded histogram).
This shows that X-ray bright groups contain a larger proportion of
galaxies that
are bigger than their counterparts in X-ray dim or undetected
groups.
\label{fig:rehist}}
\end{figure}

\subsection{Morphological classification}

The galaxies with satisfactory fits to the surface brightness
model are assigned morphological types based on their
bulge-to-total luminosity ratio, $B/T$. An ideal Elliptical galaxy
has $B/T=1$ and a late-type disk galaxy has a value of $B/T$ close
to zero.

We classify a galaxy as early-type when it has $B/T\!>\!0.5$. This
criterion was chosen to be $B/T\!>\!0.4$ by \citet{balogh02}, but the
conclusions of this study are not sensitive to such a difference in
selection. An ``Elliptical galaxy'' is one with $B/T \ge 0.95$. A
galaxy with $B/T\!<\!0.1$ is considered as a pure disk galaxy with an
exponential luminosity profile, since the bulge in this case
is too small to be modelled by the Sersic profile.

\begin{figure}
\epsfig{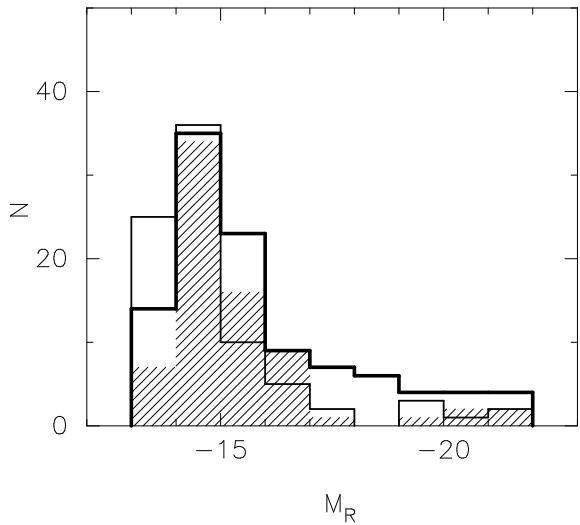}
\caption{Distribution of the absolute R magnitude
for early-type galaxies ($B/T>0.5$) in our sample.  The three
categories are the same as in Figs.~\ref{fig:nhist} \& \ref{fig:rehist}.  This
shows that X-ray bright groups contain a higher fraction of brighter
galaxies than X-ray dim or undetected groups.
\label{fig:absmaghist}}
\end{figure}

\subsection{Galaxies in X-ray bright, dim and undetected groups}

The X-ray luminosities of the parent groups in our sample are taken from
\citet{osmond03}, who measured them by fitting two-dimensional
$\beta$-profiles to ROSAT PSPC observations in the 0.5--2 keV
range, extrapolated to estimate the bolometric X-ray luminosity.
Point sources were removed from the data before luminosity
calculations and the values have then been corrected for flux lost
in this process using the surface brightness models.

Since the number of galaxies in each group is small, we obtain
statistical results by grouping galaxies according to X-ray
brightness of their parent groups. We have divided the groups to
be X-ray bright or dim according to their bolometric X-ray luminosity
being more or less than the median of the sample, $L_X=10^{41.7}$
erg/s. This luminosity refers to the X-ray luminosity of the group
plus any central galaxy that might exist. In addition, we have a
third category of group (called X-ray undetected) where there is no
discernible ``group emission'', and all the diffuse emission, if
any, can be accounted for as due to individual non-central
galaxies in the group.

There are 189, 174 and 108 galaxies (out of 316, 298 and 150) with
acceptable profile fits in the categories of X-ray bright, dim and
undetected groups respectively. Since the sample covers a wide range of
X-ray luminosity, here we seek  possible differences in the 
properties of galaxies that depend on
their local environment characterised by $L_X$.

\subsection{Variation in brightness profiles with galaxy environment}

Figure \ref{fig:nhist} shows the distribution of the Sersic parameter for
early type galaxies ($B/T\!>\!0.5$) in X-ray bright, dim and
undetected groups. There is no significant difference in the distribution of
values of the Sersic parameter $n$ among 
X-ray bright, dim and undetected groups, corroborated by
the K-S statistics.

The difference, in the distribution of the half-light radius $r_e$ of
bulges of early-type disk galaxies, between the various categories of
parent groups, is substantial. Figs.~\ref{fig:rehist} and
\ref{fig:absmaghist} show the distribution of $r_e$ and $M_R$ in the
three sub-samples in this study, indicating that X-ray bright groups
contain a larger fraction of galaxies that are bigger and more
luminous than their counterparts in X-ray dim groups. This is
confirmed by the K-S test performed on the distributions of $r_e$ and
$M_R$. The underdensity of early-types implied by the gap in the
histograms for X-ray dim and undetected groups in
Fig.~\ref{fig:absmaghist} around $M_R\!=\!-18$ is due to the dip in the LF
of these groups, discussed in further detail by \citet{miles03}.

The Sersic parameter $n$ is often found to be correlated with
the absolute magnitude of the Elliptical galaxies
\citep{young94,bing98}, which would indicate that
the Sersic parameter is an intrinsic property of a galaxy.  Here, we
find a highly scattered correlation  between $n$
and $M_R$ for Elliptical galaxies (both E and dE) in groups, with a
scatter of about 1.5~mag in $M_R$, which is much larger than the
scatter (about 0.45~mag in $M_B$) found by \citet{young94}.

\begin{figure}
\epsfig{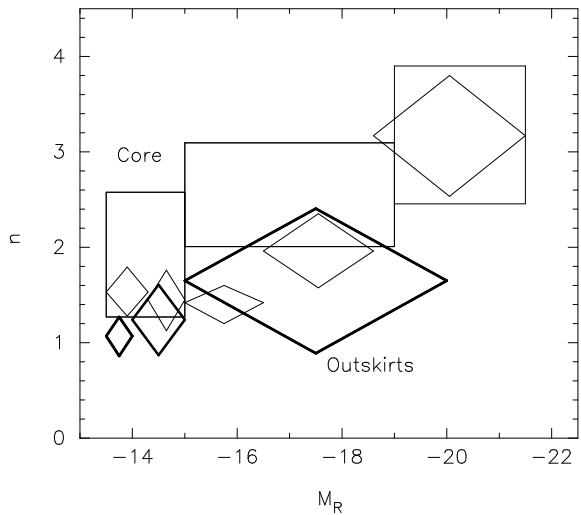}
\caption{The correlation between the Sersic parameter $n$ and the
absolute R magnitude of the Elliptical galaxies in our sample.  The
width of the diamonds represent bins chosen to include 10 galaxies in
each. The height of each diamond represents the 95\% confidence level
around the mean values of the Sersic parameter $n$ in the bin.  The
dotted and bold diamonds represent galaxies at the core ($R/R_{\rm
vir}\!<\! 0.15$) and in the outskirts ($R/R_{\rm vir}\!>\!0.6$) of the
groups, and thin diamonds represent galaxies in between.
\label{fig:nradial}}
\end{figure}

\begin{figure}
\epsfig{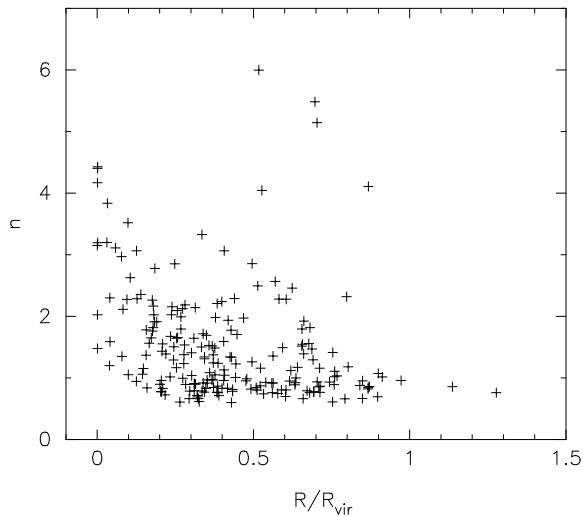}
\caption{The Sersic parameter
$n$ of each Elliptical galaxy, plotted
against the normalised projected distance from the centre of the
its parent group. This behaviour is expected if the value of $n$
in correlated with the local three-dimensional galaxy density.
\label{fig:ndistance}}
\end{figure}

In order to understand the origin of this scatter in the $n\!-\! M_R$
relation, we plot the galaxies as a function of their normalised
projected distance from the centre of their parent group
(Fig.~\ref{fig:nradial}). 
Since the distribution of the Sersic parameter
does not show a significant variation with X-ray properties of the
group, we use galaxies belonging to all three sub-samples together to
improve statistics.  Galaxies in the core of the groups ($R/R_{\rm
vir} \!<\! 0.15$) have higher values for the Sersic parameter (mean value
and 95\% confidence level of  $1.92^{+0.66}_{-0.65}$), while in the
outer regions ($R/R_{\rm vir}\!>\! 0.6$), galaxies have lower values of
$n$ and hence are less cuspy (mean value
and 95\% confidence level of  $1.07^{+0.20}_{-0.21}$).  
The systematic variation in values of
$n$ with the distance from the centre of the parent group appears to
be independent of how bright the galaxies are.

\citet{truj02} recently found a triangular distribution 
between the concentration index of galaxies (related to $n$) and the
projected two-dimensional local density of galaxies in clusters, which
is expected even if $n$ (and concentration parameter) and $M_R$ were
tightly correlated.  At higher projected densities, galaxies with both
low and high values of concentration index will be binned together due
to projection.  On the other hand, since galaxies in low-density
environments have low $n$, at lower projected densities only galaxies
with lower values of concentration index will be seen.

Fig.~\ref{fig:ndistance} shows a similar triangular distribution between
the Sersic parameter $n$ and normalised projected distance for the
galaxies in our sample.  If $n$ and $M_R$ of Elliptical galaxies were
correlated for normal and dwarf Ellipticals, Fig.~\ref{fig:ndistance}
would seem to indicate that the dwarf Ellipticals are preferentially
removed from the core of each group, but enhanced in the outer
regions.  However, from Fig~\ref{fig:nradial} it appears that the
situation is more complicated. We find faint dEs ($M_R >-15.0$), with
relatively higher values of $n$, at smaller normalised projected
distances from the centre of the groups.  This might suggest that
these high-$n$ dEs are at the core of their groups.  This strongly
suggests that environmental effects can play a key role in the shape
of the light profile (characterised by values of $n$), in galaxies and
that this parameter is not an intrinsic property of a galaxy.

This also shows that dEs, traditionally known to have near-exponential
brightness profiles, can have cuspy cores, and can exist in the
central regions of groups. These ``nucleated'' dEs have been found
to be more centrally concentrated in their distribution
in detailed studies of nearby clusters, e.g. Virgo and Fornax
\citep{ferbin94}.
It is interesting to note that such cusps are also known to exist in
starburst dEs \citep[e.g.][]{summ01}.

\section{Photometric scaling relations}

One of the essential observational ingredients of the study of
galaxy evolution is the family of scaling relations such as the
Fundamental Plane, characteristic of dynamically relaxed
Elliptical galaxies \citep{dj87,dr87}. 
Although the Fundamental
Plane for bright Elliptical galaxies is found to be scarcely
affected by environmental factors, it may not be so for low mass
Ellipticals. While \citet{nieto90} showed that dwarf Ellipticals
\footnote{Unless otherwise stated, in this paper we will refer
to Ellipticals fainter than $M_R=-18$ as dwarfs (dEs)
and those brighter as normal Ellipticals (Es).}
follow the same trend defined by the Fundamental Plane of bright
Elliptical galaxies, albeit with a larger scatter due to
structural variation, the dwarf Ellipticals in Virgo cluster
studied by \citet{peterson93} belong to a different Fundamental
Plane.

In the absence of spectroscopic measures of velocity dispersion,
which is a key parameter in constructing the Fundamental Plane,
one can study those projections of the Fundamental Plane which are
purely photometric. In this section, we will study the 
photometric characteristics of the Elliptical galaxies
in our sample (those with $B/T\!\ge \! 0.95$, as defined
in \S3.1), a subset of our sample of ``early-type galaxies''
($B/T\!\ge \! 0.5$)

\subsection{The Kormendy relation}

\begin{figure}
\epsfig{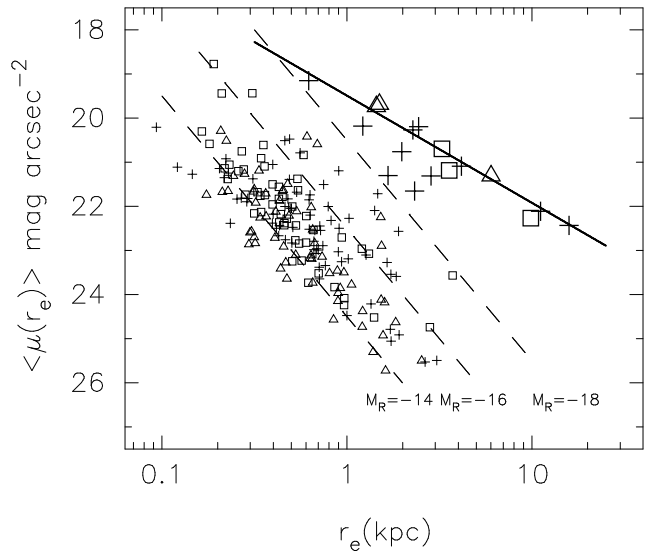}
\caption{The Kormendy relation for Elliptical galaxies in X-ray
bright (crosses), X-ray dim (triangles) and X-ray undetected groups
(squares). Large symbols are for galaxies with $M_R\!<\!-18$. The
solid line represents the KR for bright Ellipticals in the Coma
cluster with central velocity dispersion $\sigma\!>\! 200$ km/s
\citep{zieg99}.  Dashed lines represent lines of constant absolute
magnitude, our limit being $M_R=-13$.
\label{fig:kormendy}}
\end{figure}

Elliptical galaxies are known to show a strong correlation between
the mean surface brightness $\langle\mu_e\rangle$, and the
half-light radius $r_e$, known as the Kormendy relation (KR)
$\langle \mu_e\rangle =A_K\, \log r_e \, + B_K$,
with the slope $A_K$ lying in the range 2.0--3.0
\citep{korm77}. The Kormendy
relation for cluster Ellipticals shows a r.m.s. scatter of
about 0.1~dex, which would result in a 25\% error in $r_e$ and
hence in measuring distance to the galaxies.

Fig.~\ref{fig:kormendy} shows the Kormendy relation for Elliptical
galaxies in our sample, where galaxies belonging to parent groups
which are X-ray bright ($L_X\!>\! 10^{41.7}$ erg/s), X-ray dim and
X-ray undetected are plotted with different symbols.  The plot shows
two distinct populations: the ``normal'' Ellipticals ($M_R\!<\!-18$,
mean KR slope $\sim 2.3\pm0.4$) and the dwarf Ellipticals
($-18\!<\!M_R\!<\!-13$, mean KR slope $\sim 4.1\pm 0.2$).  However,
from Fig.~\ref{fig:kormendy}, it is apparent that instead of a linear KR, we
may interpret the distribution of dEs as one of high scatter,
truncated by our magnitude limit ($M_R\le -13$).

We also note that there are very few galaxies in the zone between the
two populations, which results from the significant dip in the
luminosity function of groups in the range $-16\!>\! M_R\!>\! -18$, as
seen in \citet{miles03}.

Comparing with the KR for galaxies in rich clusters, we find that the
slope of the KR for our brightest galaxies ($M_R\!<\!-19.5$)
2.75$\pm$0.16, compared to 2.43$\pm$0.15 for a similar sample in the
Coma cluster (\citealt{zieg99}, plotted as the solid line in
Fig.~\ref{fig:kormendy}).

Extending the KR plot of the Coma cluster to fainter galaxies
\citep[e.g.][]{graham03}, Ellipticals in the magnitude range
$-15\!>\!M_B \!>\!-18$ are found to follow a trend, at almost constant
$r_e$, quite distinct from that of the brighter galaxies, the latter
having a slope similar to that found in our sample. Our sample of
dwarfs extends to fainter $M_R \approx -13$ smaller ($r_e \approx
100$~pc) galaxies, and shows the presence of substantial scatter in
both axes.  Some contribution to this large scatter may also arise
from dependencies on local environment, since we are merging galaxies
from 16 groups in a single plot. Our conclusion (in agreement with
\citet{graham03}) is that the Kormendy relation in itself does {\it
not} provide strong evidence for any fundamental difference in
structure between normal and dwarf Ellipticals.

\begin{figure*}
\epsfig{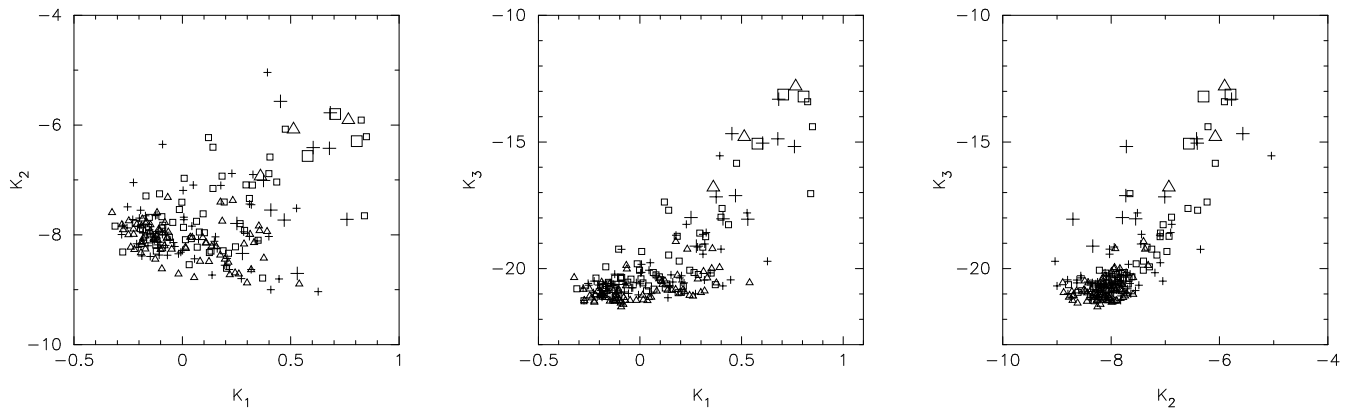}
\caption{The face-on and two edge-on views of the Photometric Plane 
for Ellipticals in our sample. The symbols have the same meaning
as in Fig.\ref{fig:kormendy}. the principal axes $K_1$, $K_2$ and $K_3$
are defined in \S4.2.
\label{fig:photplane}}
\end{figure*}

\subsection{The Photometric Plane}

Recent developments in the study of the structure of galaxies and the
modelling of brightness profiles, in particular with the use of the
more general Sersic model, provides a unified description of the
photometric properties of normal (E) and dwarf
Ellipticals (dE), traditionally seen as two distinct families
\citep[e.g.][]{korm85}.  The discontinuity between these two extreme
exponential ($n\sim 1$, dEs) and \dev\ profiles ($n\sim 4$, Es) is now
filled by both populations possessing a large range of values $1\!<\!
n\!<\! 6$. The
bulges of disk galaxies share the same Photometric Plane with 
Ellipticals.

The Photometric Plane \citep[PP,][]{habib00a}
is a bivariate relation that links 
only photometric parameters, obtained by fitting a Sersic model
to a galaxy image,
$\log n=A\, \log r_e \, + B\, \mu_0$.
Our simulations show that fitting galaxy brightness distributions with
unsuitable models (e.g. the \dev\ model to a galaxy where $n\ne 4$),
can result in a significant variation in the values of structural
parameters.  For example, assuming $n=4$ in the case of a $n=2$
galaxy can result in a change in the value of half-light radius 
by a factor of three. Thus it is important to allow for a free
parameter $n$ to account for the cusp and core in Ellipticals or
bulges of disk galaxies.

Most scaling relations, e.g. the Kormendy relation or the
Fundamental Plane, are  insensitive to these variations, since
the three parameters of the Sersic model ($\mu_0$, $n$, $r_e$) are
correlated and their combinations used in these scaling relations are
robust \citep{bertin02}. The Photometric Plane, on the other hand, 
allows us to explore the entire range of
these  parameters for Elliptical galaxies, and thus can be used
more effectively as a tool to examine the connection between
structural parameters and the evolutionary history of galaxies.

The Photometric Plane for all Ellipticals in our
sample is given by
\begin{eqnarray}
\log n = & (0.21 \pm 0.09)\, \log r_e &\nonumber\\
         & - (0.074 \pm 0.013)\, \mu_0 &+ (1.7 \pm 0.3).
\label{eq:ppeq}
\end{eqnarray}
This equation is similar to that found for the bright
Elliptical galaxies in the $K$-band study of 42 Coma galaxies
\citep{habib00a}. However, the scatter in $n$ here is 0.13 $dex$, much
larger than the scatter found for the Coma cluster sample, which is
0.05 $dex$, which can be attributed to the fact that our sample is
dominated by dEs. To verify this, we examine the Ellipticals with
$M_R< -18$ in X-ray bright groups (comparable to the Coma sample), to
find that the scatter about the Photometric Plane for Es is only 0.06
$dex$ in $\log n$, compared to 0.10 $dex$ for dEs.  The fact that the
scatter around the the Photometric Plane of normal Es is lower than
that of dEs indicates that Es and dEs are structurally different
systems even though their $n$ values might be similar.

We plot the principal axes ($K_1$, $K_2$, $K_3$) of the Photometric
Plane against each other in Fig.~{\ref{fig:photplane}}, where $K_1=0.99
\log n+0.17 \log r_e$, $K_2=0.16 \log n-0.92 \log r_e -0.38 \mu_0$ and
$K_3=-0.06 \log n+0.37 \log r_e -0.93 \mu_0$ (directly comparable to the
similar plot for Coma Ellipticals in 
\citealt{habib00a}). 
Since this sample has a fainter magnitude limit, it probes the PP for
dwarfs more extensively, and demonstrates that the PP for the dEs has
a is strikingly low scatter (in contrast, for example, to their
scatter in the Kormendy plot).  This strongly suggests that dEs might
be a more homogeneous class than is normally thought, and follow
well-defined structural trends.

To explore a more direct connection with the underlying physical
changes in galaxy structure between Es and dEs, we turn to a slightly
different version of these correlations in the next section.

\subsection{The Specific Entropy Plane}

\citet{mar01}, in their study of cluster Ellipticals, showed that the
Photometric Plane is close to a plane of constant specific entropy
of galaxies, originally introduced by \citet{lgm99}. Although 
self-gravitating systems like galaxies do not posess stable states 
of absolute maximum entropy, the evolution of entropy via two-body 
relaxation will be a slow process, and \citet{mar00} argue that the
violent relaxation process which occurs during galaxy formation may
establish a quasi-equilibrium state of constant specific entropy,
and they find evidence that both simulated and observed galaxies do
indeed show approximately constant entropy per unit mass. In the case
of observed galaxies, the properties of the Photometric Plane result
from this approximate constancy of specific entropy.

\begin{figure}
\epsfig{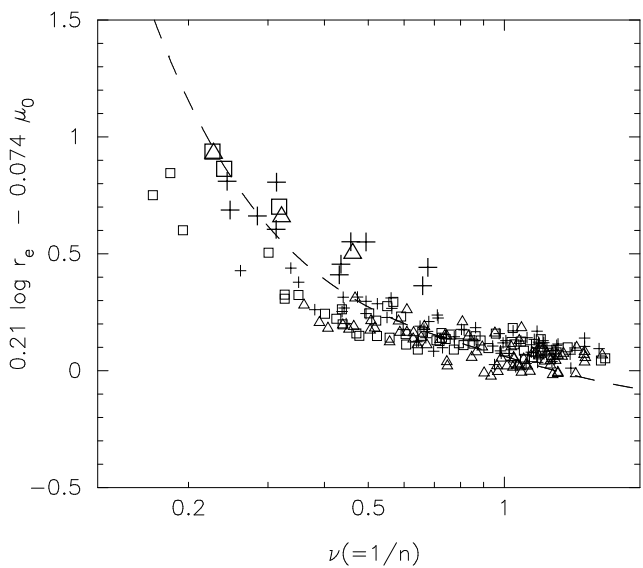}
\caption{The Photometric and Specific Entropy Planes of Elliptical 
galaxies in X-ray bright (crosses), X-ray dim (triangles) and X-ray
undetected groups (squares). Large symbols are for galaxies with
$M_R\!<\!-18$.  The dashed curve is the best match Specific Entropy
Plane with $s_0=-25$.
\label{fig:entropy}}
\end{figure}

On examination of the surfaces of constant entropy in detail, it is
found that they actually exhibit some curvature when represented in
the coordinate system of the PP. It can be seen, from the two edge-on
views of our plane in Fig.~\ref{fig:photplane}, that there are clear
indications of curvature in the distribution of the dwarfs. This
curvature, rather than stochastic scatter, accounts for much of the
larger variance about the plane for dEs (compared to nornal Es)
discussed in section~4.2. Could this curvature be indicating that
dwarfs really follow a surface of constant specific entropy? In
Fig.~\ref{fig:entropy} we plot a representation of the Photometric Plane
in terms of $1/n$, for our sample, using Eq.~(\ref{eq:ppeq})
above, and overlay a line of constant specific entropy. The latter has
been calculated using Eqs.~(23) \& (24) of \citet{mar01}, in
conjunction with the relevant relations from their Appendix A. Here,
we have only adjusted the value of the specific entropy parameter
($s_0$) to find the curve which best matches our data.

The curvature seen in the distribution of dwarfs in
Fig.~\ref{fig:entropy} is similar to that the constant $s_0$ surface,
although the shape does not match in detail. It is also apparent that
the normal Es fall above the dwarfs in the Figure, which corresponds
to the direction of increasing $s_0$. The dashed line corresponds to a
specific entropy Plane with $s_0=-25$ for our sample of Elliptical
galaxies dominated by dEs.

\section{Discussion and Conclusions}

This study confirms earlier results that massive Ellipticals show
higher values of the Sersic parameter $n$, {\it i.e.}  a cuspier
surface brightness profile at their centre \citep{young94}. From the
large scatter in the plot of absolute R magnitude and $n$
(Fig.~\ref{fig:nradial}), we conclude that the shape of the radial
profile, represented by the value of $n$, is not uniquely related to
the intrinsic luminosity of the galaxy. Furthermore, the dependence of
$n$ on the location of galaxies within their groups
(Fig.~\ref{fig:nradial}) shows that environmental dependence is
responsible for part of this scatter. Although we find no significant
difference between the values of $n$ in groups of different X-ray
luminosity, we do find that for both dEs and Es (though not the
most luminous Es), surface brightness profiles tend to be cuspier
(higher $n$) for galaxies located in the cores of groups. One
explanation for this correlation could be the effects of
interaction-induced star formation in triggering nuclear starbursts
within more centrally located galaxies.  

Normal Ellipticals in groups are found to yield similar Photometric
Plane and Kormendy relation parameters to those in rich clusters.
There is no difference in the Kormendy relation or Photometric Plane
between Ellipticals belonging to the the different classes of X-ray
bright, faint or undetected groups.  Furthermore, the Kormendy
relation is found to be of little value in distinguishing structural
differences between dEs and Es, since the KR slope found for dwarfs is
dominated by selection effects, most notably the magnitude limit of
the sample. The sparse region in the KR plot, separating dEs and Es,
is easily interpereted as resulting from the dip seen in the galaxy
luminosity function at $M_R\sim -18$, and discussed in more detail in
\citet{miles03}.

Dwarf and normal Ellipticals
are distinguished much more effectively in terms of
their location in the Photometric Plane in the space of $n$, $r_e$ and
$\mu_0$. The dwarfs inhabit a non-flat surface in this space, with a
curvature somewhat akin to a surface of constant specific entropy, as
defined by the equations of \citet{mar01}, though the match is only
approximate. Even if galaxies do have constant specific entropy, a
degree of mismatch between observation and the analytical expressions
of \citet{mar01} is not surprising, given the numerous assumptions
(e.g. mass-traces-light, and isotropic velocities) which have been
made in deriving the equations of the isentropic surfaces. The
well-defined nature of the locus followed by these dwarfs is a strong
indicator of their structural coherence, and the larger scatter that
they exhibit about their Photometric Plane is largely due to the
curvature of the surface they inhabit, rather than to a large random
scatter in their properties.  Normal galaxies, on the other hand, are
offset relative to the dwarf locus, in a direction corresponding to
higher specific entropy. In this context, it is interesting that
\citet{mar00} find, from their simulations, that merging tends to
increase specific entropy. Hence the offset may indicate that normal
Ellipticals,
which fall above the dip seen in the galaxy luminosity function, have
undergone mergers, whilst dwarfs have not. This supports the idea
\citep{miles03} that merging may have played a role in generating
the dip in the luminosity function, by converting middle-ranking
galaxies into larger ones, whilst dwarfs are largely unaffected due to
their smaller merger cross-sections.

\section*{Acknowledgments}

We would like to thank Ale Terlevich for help in observation and
basic data reduction, John Osmond for making the results of his
analysis available for us and Yogesh Wadadekar for the use of his 2D
decomposition code.

\bsp

\label{lastpage}
\end{document}